# Habitability in Different Milky Way Stellar Environments: a Stellar Interaction Dynamical Approach

Juan J. Jiménez-Torres Instituto de Astronomía, Universidad Nacional Autónoma de México
Bárbara Pichardo Instituto de Astronomía, Universidad Nacional Autónoma de México George Lake Department of Theoretical Physics, University of Zürich Antígona Segura Instituto de Ciencias Nucleares, Universidad Nacional Autónoma de México

To whom correspondence should be directed:

Juan J. Jiménez-Torres

Hamburger Sternwarte

Gojenbergsweg 112, 21029 Hamburg, Germany

E-mail: jtorres@hs.uni-hamburg.de

**Running title:** Milky Way Habitability: A Dynamical Approach

**Abstract**

Every Galactic environment is characterized by a stellar density and a velocity dispersion. With this information from literature, we simulated flyby encounters for several Galactic regions, numerically calculating stellar trajectories as well as orbits for particles in disks; our aim was to understand the effect of typical stellar flybys on planetary (debris) disks in the Milky Way Galaxy.

For the Solar neighborhood, we examined nearby stars with known distance, proper motions, and radial velocities. We found occurrence of a disturbing impact to the Solar planetary disk within the next 8 Myr to be highly unlikely; perturbations to the Oort cloud seem unlikely as well. Current knowledge of the full phase space of stars in the Solar neighborhood, however, is rather poor, and thus we cannot rule out the existence of a star that is more likely to approach than those for which we have complete kinematic information. We studied the effect of stellar encounters on planetary orbits within the habitable zones of stars in more crowded stellar environments, such as stellar clusters. We found that in open clusters habitable zones are not readily disrupted; this is true if they evaporate in less than $10^8$ years. For older clusters the results may not be the same. We specifically studied the case of Messier 67, one of the oldest open clusters known, and show the effect of this environment on debris disks. We also considered the conditions in globular clusters, the Galactic nucleus, and the Galactic bulge-bar. We calculated the probability of whether Oort clouds exist in these Galactic environments.

1. Introduction

Debris disks are extremely fragile entities that can be dynamically perturbed by a number of processes, for example, planet-planet interactions (Batygin, Morbidelli & Tsiganis 2011), interactions between giant planets of similar size (Raymond et al. 2011, 2012), secular perturbations by giant planets (Mustill and Wyatt, 2009), migrating planets (Walsh et al. 2011), the presence of a stellar companion (Paardekooper et al. 2012, Thébault 2012, Kaib et al. 2011), or gravitational interactions with other stars (Jiménez-Torres and Pichardo 2008, Lestrade et al. 2012). The most violent of these processes are stellar interactions. This type of encounter can readily and rapidly change orbital parameters of planets and minor bodies (Jiménez-Torres and Pichardo 2008). The other processes affect dynamics at much larger timescales and may be more important in regions where stellar density is smaller (solar neighborhood or young stellar clusters), and the number of encounters is less probable. Nevertheless, any single close stellar encounter can change disc orbital dynamics dramatically, in such a way that its effect might not be erased by other typical processes like planetary interactions (one example is the Kuiper belt). In this work, we focus on the study of the effect of stellar encounters, characteristic of different Galactic environments, on debris disks (planets, cometary nuclei, etc.) to determine the dynamical stability and habitability potential of planets around stars.

In previous work, we studied the gravitational effects of stellar interactions on the orbital dynamics of debris disks (Jiménez-Torres *et al.* 2011). As an extension of that work, we now examine the possibilities for habitability of planetary systems in different regions of the Galaxy, based on disks dynamical studies. To construct the parameters of stellar encounters, we divided the Galaxy into different regions or "environments": the Solar neighborhood, open clusters (including the birth cloud of the Sun), globular clusters (the case of Messier 13), the Galactic bulge-bar, and the Nucleus. Every environment is characterized by a stellar density and a velocity dispersion that allow for characterization of average stellar encounters. We analyzed the dynamical effect of one stellar encounter on a model of a planetary (debris) disk. Potential habitability of planetary systems could be disrupted by such encounters in three ways: 1) perturbing the protoplanetary disk and preventing the formation of planets in the habitable zone, 2) perturbing the orbital parameters of planets in the habitable zone and rendering the planetary surface of those planets hostile for life or producing close encounters of external perturbed planets with the inner terrestrial planets (Zakamska and Tremaine 2004, Haghighipour and

Raymond 2007), both processes produced by highly eccentric planetary orbits, and 3) perturbing Oort cloud-like objects and potentially enhancing the flux of comets towards the inner planetary system.

Our dynamical approach identifies the most likely Galactic environments where stars can form and keep their planets and comets, which is in turn related to the possibility of life in these systems.

Since the dynamical evaporation from its birth cluster, our Solar System has not sustained close interaction with any star of the Solar neighborhood. Furthermore, even if we consider that the Sun has migrated radially (Sellwood & Binney 2002; Roskar et al. 2012, and references therein) such that it was closer to the Galactic center in the past where the density of the environment was greater, the estimation for the radial migration of the Sun is about 2 kpc. Even if the Sun migrated 4 kpc (more than models can explain), that is, if the Sun was born at 4.5 kpc, this is more than 1.5 times the radial scale of the Milky Way (~2.5 kpc). From that distance to the outskirts of the Galaxy, the density is so low that migration would not have affected the present results. However, there is some evidence that suggests a stellar interaction with the solar system occurred at some point in the past, most likely in its birth cluster (Goswami and Vanhala, 2000; Hester *et al.* 2004; Ida *et al.* 2000; Kobayashi and Ida 2001; Looney *et al.* 2006; Meyer and Clayton 2000; Wadhwa *et al.* 2007). Among the most important pieces of evidence of interaction is the dynamical heating of the Kuiper belt, that is, cometary nuclei in the Kuiper belt have larger eccentricities and inclinations than can be explained by secular resonances with Neptune or collisions among the Kuiper belt objects. This dynamical heating may have been the result of external excitation (Ida *et al.* 2000; Luu and Jewitt, 2002). Also, the Kuiper belt shows a significant drop-off at around 50 AU (Melita *et al.* 2002), and beyond the Kuiper belt, there exist objects like Sedna, with large eccentricities and inclinations that cannot be explained by planetary scattering (Brown *et al.* 2004). Other model studies have attempted to explain the ring of cometary nucleii or *scattered Kuiper belt objects,* which surround the solar system beyond the orbit of Pluto. These objects present high *e* and *I*, invoking Earth-sized bodies that are assumed to have existed at some time in the formation stage of the Solar Sytem and ejected later (Morbidelli and Valsecchi 1997; Petit, Morbidelli, and Valsecchi 1999). Partial trapping by sweeping of the 2:1 resonance might have also excited the Kuiper belt at about 45 AU (Hahn and Malhotra 1999). From all the models known to explain the characteristics of the Kuiper belt, the stellar flyby model has been one of the most succesful (Ida *et al.* 2000; Kenyon and Bromley 2004; Levison et al. 2004). All these characteristics point to the possibility that the Solar System sustained an ancient encounter. This raises the question, How do stellar environments influence the formation and evolution of Oort-like clouds, Kuiper-like belts, and planets?

In dense environments, the number of interactions between stars may be high enough to partially or totally destroy planetary systems, that is, break their gravitational link with the host star. In globular clusters, for example, where transient searches for planets have a very low detection probability, planets may exist around main-sequence stars, although probably at small numbers because of the low metallicity (Soker and Hershenhorn, 2007). Indeed, bound planets have been discovered (Sigurdsson et al. 2003), such as those in the system PSR 1257 + 12 (Wolszczan and Frail, 1992). Free-floating planetary-mass objects have been detected as well (Sahu *et al.* 2001), and these planets might be evidence of frequent destructive stellar encounters. This stellar environment is characterized by high densities, slow velocity dispersions, and the longest lifetimes, all of which reinforce the destructive effect on disks. In this manner, disks are dynamically unstable on much shorter time scales compared to the typical age of globular clusters (Sigurdsson, 1992).

For the Galactic bulge, there are a few surveys, such as that of Sahu *et al.* (2006), that show the results from a planetary transit search performed in a rich stellar field towards the bulge. These authors discovered 16 candidates with orbital periods between 0.4 and 4.2 days, five of them orbiting stars with masses in the range 0.44-0.75 $M_\odot$.

This paper is organized as follows. In Section 2, we describe how planetary habitability may depend on orbital eccentricity. Section 3 shows the method and numerical implementation. In Section 4, we present our results for several regions: the Solar Neighborhood, the birth cloud of the Sun, the open cluster Messier 67, the globular cluster Messier 13, the Galactic Bulge-Bar and the Galactic Nucleus. Conclusions are presented in Section 5.

2. The eccentric habitable zone

Our future search of habitable exoplanets is based on the concept of surface habitability, that is,. rocky planets with the environmental conditions required to support liquid water on its surface (Kasting *et al.* 1993; Gaidos and Selsis, 2007; Kaltenegger *et al.* 2010). One requirement is the location of the planet around its host star. The habitable zone (HZ) is defined as the annulus around a main sequence star where a planet with an atmosphere can support liquid water for a given time (Kasting *et al.* 1993).

The most commonly used limits for the habitable zone are those calculated by Kasting et al. (1993), where orbital eccentricity was not included. Williams and Pollard (2002) showed that planets with elliptical orbits might be suitable for life as long as they receive an average stellar flux similar to the nearly constant Solar flux received by Earth around the Sun.

The time-averaged flux over an eccentric orbit is given by

$$\langle Q \rangle = \frac{L}{4\pi a^2 \left(1-e^2\right)^{1/2}} \tag{1}$$

where $L$ is the host star luminosity (Williams and Pollard, 2002). Barnes et al. (2008) used this averaged flux and the limits of the habitable zone calculated by Kasting et al. (1993) to define the *eccentric habitable zone* limits (Eq. 8 and 9 in Barnes et al. 2008). Fig. 1 shows the boundaries for the eccentric habitable zone for a solar mass star ($1L_\odot$, $T_{eff}$=5700 K) by using the two limits calculated for each boundary of the habitable zone (Table I in Kasting et al. 1993). It is important to recall that the limits of the HZ are calculated by climate models for a given atmosphere. Cloud coverage, for example, is critical for setting the limits of the HZ. The cloudless model used by Kasting et al. (1993) calculated the boundaries between 0.84 AU and 1.7 AU, while a planet with 100% cloud coverage remains habitable between 0.46 AU and 2.4

AU (Selsis et al. 2007 and references therein). When eccentricity is included, a faster rotational period of the planet extends the limit of the outer HZ for eccentricities larger than ~ 0.65 (Dressing et al. 2010). Other parameters can be taken into account for determining planetary habitability like ocean-land distribution (e.g., Dressing et al. 2010) or tidal heating (e.g., Barnes et al. 2008). For the purposes of this work and to be as general as possible, we used the limits presented in Fig. 1 to analyze the effect of a stellar flyby in the habitable planets around stars located in different Galactic environments.

3. Numerical Implementation: The Stellar Encounter Code and the Employed Parameters

We have built a numerical code that simulates a three dimensional 100 AU disk, under the influence of a stellar hyperbolic encounter. The disk is made of test particles evenly distributed in phase. Test particles are affected by the gravitational forces of both the central (host) star and the flyby star, and the equations of motion are solved in the non-inertial frame of reference centered on the host star. We calculated the orbital parameters, such as eccentricities and inclinations of particles after the stellar impact, once the flyby star is far enough to make its gravitational influence negligible. We employed the adaptive Bulirsh-Stoer (Press *et al.* 1992) integrator that provides a maximum relative error, much before and after the encounter of less than $10^{-13}$ in the energy and angular momentum integrals. Sampling of the orbits goes like $a \propto N^{3/2}$, where $a$ is the initial radius of a given orbit and $N$ is the number of orbit. Fig. 2 (taken from Jiménez-Torres *et al.* 2011) shows a scheme of the relevant parameters used in the code for a stellar encounter. The dark disc at the center of the system represents the debris disc, the grey sphere radius represents the maximum approach distance of the flyby, and the small disc is tangent to the sphere at the point of minimum distance. The flyby trajectory angles are $\phi$, the azimuthal angle with respect to the disk, which ranges from 0° to 360°; the polar angle with respect to the disk, from -90° to 90°; and $\alpha$, the angle between the flyby plane orbit and the symmetry axis of the planetary disk, which goes from 0° to 360°.

Regarding the flyby attack angle, $\phi$, which is due to the azimuthal symmetry of the system initial condition, this entrance angle is indistinct; we took for our experiments $\phi = 0°$. In the application presented in low density environments of the Galaxy, the effects of the exact direction of entrance of the flyby are (of?) no importance. For example, in the current Solar neighborhood, stars will not come close enough for the angle to become important, that is, in the majority of cases the effect of encounters at such large distances would be negligible, even for objects in the Oort cloud. In the case of the birth cloud of the Sun, star approaches are slightly more likely to be closer; in this case the flyby direction of entrance could be important; however, disks orientation with respect to flyby orbits are rather random, that is, in general there is no preferential direction of entrance for the encounters that could be used to calculate the most probable trajectories.

Because we are talking about a single encounter, we chose a general interaction, that is, 45° for $\theta$ and $\alpha$, that should produce an intermediate effect in each angle that defines the flyby orbit. This means that the flyby orbit will enter at 45° with respect to the plane of the disk and 45° with respect to the disk axis. It is worth noting that the results are nearly insensitive to changes of the angle $\alpha$, while changes in the polar angle produce moderately different results, being larger for angles closer to 0° (parallel to the plane of the disk); this angle acts as a multiplicative parameter, which means that the behavior of the resultant particle disk inclinations and eccentricities is similar for a given encounter, but multiplied by a small factor that increases as the polar angle decreases. In the polar angle case, we simply used a single value in the middle (45 deg), but the behavior would be similar for other angles, especially in low dense environments. In the same manner, we are generally presenting for every environment interactions of one solar-mass (unless the contrary is specified) in the case of the host star, because we are trying to represent habitable solar system-like planets (the only life form we know). This represents a "conservative" limit since we are assuming host stars that are slightly more massive than the average in most environments, which produce stronger discs (from the gravitational point of view) that are more capable of retaining planets. For purposes of simplicity, in this first approximation we assumed one single hit by a solar mass flyby star, perfectly plausible in all environments. Although some environments, such as globular clusters are crowded enough to allow several encounters in the life of the star in the cluster, the first encounter excerts the most destructive effect. . In a future paper, a comprehensive study of the dynamics of cluster environments, with multiple encounters with average masses in every environment, will be presented. In this present paper, we present a first approximation from the dynamical point of view of all the environments, which provides instructive insight into the survival of discs in different stellar systems of the Milky Way.

4. Results

With the restrictions posed by planetary eccentricity on habitability, we examined those stellar regions in the Galaxy that are suitable for life with regard to the stellar interaction dynamics. Planetary systems may become highly eccentric due to stellar interactions typical of almost every Galactic environment. We have endeavored to understand, at a first approximation, which regions of the Galaxy are capable of mantaining unperturbed planetary disks. For this purpose, we calculated the effect of stellar encounters typical of different Galactic environments on a 100 AU planetary disk. We present in Figure 3 a log-log plot of density *vs.* velocity dispersion and show show with level curves the number of encounters for a given combination of these two parameters in different Galactic environments. The Galactic environments are marked with elliptical regions approximately where they correspond according to their physical characteristics. Straight lines represent the number of encounters (eq. 4), given a density and velocity dispersion for a total integration time $T_e$ of 5 Gyr (for all environments). All environments included have existed for the integration times we employed ( most of the environments existed for longer times than the integration times used), except for young clusters (they live bounded about $10^8$ years), whose environment is so rarified that the number of encounters is almost the same in the total integration time $T_e$ employed. The green shadow covers the Galactic regions that sustained in its history less than one stellar encounter. These regions are potentially habitable from the stellar encounter dynamics point of view. In the case of globular clusters, we plotted the density and velocity dispersion of their typical central regions; however, stars in the outskirts of these clusters have densities and relative velocities suitable for keeping planetary disks stable and potentially for life (dynamically talking). It must be taken into consideration, however, that a large fraction of stars in these type of clusters have low orbital angular momentum, that is, they present periodic radial excursions to the central, much denser parts of the cluster, which increases dramatically the possibility of disruption. Central parts of globular clusters and central regions of the Galaxy are the most affected by repeated encounters. Below, we consider several regions of the Galaxy with regard to the plausibility of life within them.

### 4.1. The Solar Neighborhood

What is the risk that Earth to be altered or rendered uninhabitable by a body from outer space? Normally, this question carries the context of a direct hit by small bodies of the Solar System. Here, we ask whether there is a risk due to passing stars that have the potential to alter the delicate equilibrium of our planetary disk and change the orbit of Earth, Jupiter, the Kuiper belt, the asteroid belt, or any body of the Solar System that would affect life on Earth. Studies like this are not new (Bobylev, 2010; García-Sánchez *et al.* 2001, 1999; Jiménez-Torres *et al.* 2011; Matthews, 1994); however, missions like GAIA (Lindegren *et al.* 2008; Perryman, 2005; Mignard, 2005) and RAVE (Smith *et al.* 2007; Steinmetz *et al.* 2006; Zwitter *et al.* 2008; Steinmetz, 2003) are designed to measure stellar radial velocities, parallaxes, and proper motions, and may introduce many more stars to the close approach samples presented in literature to date.

We compiled 1167 stars of the Solar Neighborhood from different catalogs (Hipparcos, Nexxus 2, Simbad) to gather all the information needed such as parallax, proper motions, radial velocities, and equatorial coordinates. A table with the data, including the miss distance of each one of the 1167 stars with the Sun and the time of maximum approach, was presented by Jiménez-Torres *et al.* (2011). In that work, the authors computed the orbits of all the stars in the sample in a global Galactic potential observationally motivated (Pichardo *et al.* 2003, 2004).

From the objects in this sample, the star Gliese 710 will approach closest to the Sun. This 0.6 $M_\odot$ star is currently located at 19.3 pc (*García-Sánchez et al.* 1999). We obtained a miss distance Sun-Gliese 710 of 0.34 pc (70,000 AU), $1.36 \times 10^6$ yr in the future. Fig. 4 is a close up of Fig. 2 from the Jiménez-Torres *et al.* (2011) study. The figure shows miss distance versus time of maximum approach for the stars in the Solar neighborhood whose closest approaches to the Sun are shorter than 3 pc.

Even the closest approach will have a negligible effect on our planetary system because of the enormous miss distance (0.34 pc ~ 70,000 AU). It takes a 300 AU encounter with a solar mass star to create a slight but clear perturbation in the outskirts of a 100 AU planetary disk (Jiménez-Torres *et al.* 2011). The case of Gliese 710, which will be the closest known approach to the Solar system in the near future, is interesting albeit the star will only cross the outer limit of the Oort Cloud (~70,000 AU) and thus the effects on the Oort cloud will be slight although not negligible (Bobylev, 2010; García-Sánchez *et al.* 2001, 1999; Hills, 1981; Matthews, 1994; Weissman, 1996).

### 4.1.2. The Oort Cloud in the Solar Neighborhood

An estimation for which the impulse approximation was used predicts that, during a stellar encounter, the Sun's velocity would change by $\delta v = (2\,G\,M_s) / (q_s\,v_\infty)$, where $M_s$ is the flyby star mass, $q_s$ is the miss distance, and $v_\infty$ is the Sun-star relative velocity. We take the approximate values for Gliese (710, $M_s$ = 0.6 $M_\odot$, $q_s$ = 0.34 pc, and $v_\infty$ = 13.4 km/s) and obtain 5.8 × $10^{-4}$ km/s for the change in velocity. This is a small value compared to the typical orbital velocity of the Oort cloud objects, which is 0.2 km/s. This implies that the gravitational effect by Gliese 710 would not have a destructive effect on the Oort cloud. However, as García-Sánchez *et al.* (2001) showed, this star has the potential to send an enhanced flux of comets toward the inner Solar System. This would increase the flux to roughly one new comet per year sent toward the inner Solar System. In these terms, the risk of an Earth impact is negligible.

### 4.2. Birth Cloud of the Sun

In star formation regions such as the Sun's putative birth location inside a dense cluster, stellar densities are high enough to produce stellar encounters within 200 AU before the dissolution of the stellar cluster (Laughlin & Adams 1998; Adams 2010). The stellar densities in this regions would produce a 20% probability of a stellar encounter within 200 AU before the star cluster dissolves in approximately $10^8$ yr (Shu *et al.* 1987; Carpenter, 2000; Lada and Lada, 2003; Looney *et al.* 2006; de la Fuente Marcos and de la Fuente Marcos, 1997, Laughlin and Adams, 1998, Hurley and Shara, 2002, Pfahl & Muterspaugh, 2006, Spurzem *et al.* 2009; Binney and Tremaine, 2008). The typical velocity dispersion in these Galactic environments is very low, about 1-3 km/s, which increases the time for interactions and therefore the gravitational effects of stellar encounters on disks. Still, neither the encounter distance nor the slow velocity seem to be enough to alter inner planetary orbits in this environment.

However, such an encounter could provide an explanation for features in our cometary systems (Ida *et al.* 2000, Kobayashi *et al.* 2001; Kobayashi *et al.* 2005). The classic Kuiper belt, for example, has an abrupt edge at 50 AU. It is dynamically excited (eccentricities up to 0.4), and it has objects with high eccentricities and apocenter distances that cannot be readily explained considering interactions between the components of the Solar System. All these features are used as restrictions to calculate the physical and orbital characteristics of the likely Solar System disturber.

For this purpose, Jiménez-Torres *et al.* (2011) ran numerical simulations with a Monte Carlo scheme for initial conditions with different impact parameters and relative velocities. In Fig. 5, we present one of these experiments with an approach distance Sun-star of 150 AU and relative velocity of 1 km/s. The figure shows the eccentricity (upper left panel) and inclination (upper right panel), pericenter and apocenter distances (lower left and right panels), all versus semimajor axis. As reference, we have included the observed Solar system; resonant and classic Kuiper belt objects are in pink triangles and the scattered objects (including the Centaurs at radii less that 30 AU) are in cyan crosses.

In this experiment, we reproduced approximately the observed Kuiper belt characteristics with a single flyby interaction. This is the most accepted mechanism to explain the Kuiper belt edge (Ida *et al.* 2000) and dynamical heating. This experiment, as depicted in Fig. 5, produced eccentricities up to 0.4 for particles located between 35 and 45 AU, and the inner regions in the interval (up to 5 AU) were not disrupted. A planet inside this limit would preserve its circular stable orbit.

### 4.2.1. The Oort Cloud in the Birth Cloud of the Sun

Using the impulse approximation, we estimated whether the Oort cloud could survive flyby encounters in this Galactic environment. Typical escape velocities of Oort cloud objects are 0.25 km s$^{-1}$. Therefore, a minimum condition under which the Oort cloud would not be stripped out would be such that the particles do not reach the escape limit velocity, $v < 0.25$ km s$^{-1}$ or in other words, it would mean that, to keep the Oort cloud in a star-star encounter, , the (Oort cloud) host star could not change its velocity violently far beyond this limit:

$$v_e = 0.25 \text{ km s}^{-1} > \frac{2GM_s}{q_s v_\infty} \qquad (2)$$

that is, with $G = 889.105$ AU M$^{-1}$ (km s$^{-1}$)$^2$, equation 2 takes the form,

$$q_s v_\infty > 7113 M_s \qquad (3)$$

where $[q_s]$ = AU, $[v_\infty]$ = km s$^{-1}$ and $[M_s]$ = M$_\odot$.

Applying this to typical parameters in the birth cloud of the Sun, , 200 [AU] × 3 [km/s] = 600 < 7113. This means that, in the original condition of the birth cloud of the Sun, the Oort cloud should have been readily stripped out. Because the Oort cloud is clearly still there, judging from the existence of long period comets, this takes researchers to the conclusion that the time required for the Oort cloud to form was larger than the duration of the flyby approach that heated and truncated the Kuiper belt. In this manner, the mechanism that formed the Oort cloud (probably expulsion of planetesimals from the inner Solar System by the giant planets) seems to have taken about 10$^9$ years. This is a long time compared to the maximum time when the flyby should have taken place, and compared to the cluster dissociation time, which is about 10$^8$ years (Duncan *et al.* 1987; *Fernández,* 1997; Dones *et al.* 2004; Levison *et al.* 2004).

## 4.3. The Open Cluster Messier 67

In general, over the lifetime of a typical open cluster no more than one close encounter (of less than 200 AU) per star is likely (Malmberg *et al.* 2007), except for very low angular momentum stellar orbits that will pass close to the center of the cluster where density is higher and for long lasting clusters, such as Messier 67 (M67).

### 4.3.1. The current Messier 67

Messier 67, which is located in the constellation of Cancer, is an atypical open cluster. It is the oldest open cluster known, and it is still gravitationally bound. Its stars are approximately the same age as the Sun (4 billion years) and have similar metal abundances (Biazoo *et al.* 2009A,

2009b; Pasquini *et al.* 2008). We consider in this section the effect of a stellar encounter in this environment, on a 100 AU planetary system.

Considering the density and lifetime of this environment, we approximated the stellar dynamics of M67 as it is in the present and as it was in the distant past. For this purpose, we calculate the mean free path, which is given by λ=1/σn, where $\sigma$ and *n* are the cross section and stellar density per number, respectively. Dividing the mean free path by the velocity dispersion ν, we obtain a characteristic time *T*, on which an encounter occurs. We can calculate the number of encounters by dividing *T* by a given total time $T_e$ (for example the age of the cluster). On the other hand, we know that a flyby must occur within a 200 AU cross section to produce a noticeable perturbation on a 100 AU disk. We define the cross section as $\sigma = \pi (2R)^2$ where *R* is the radius of the planetary system (100 AU). In this manner, the number of encounters is given by

$$N_e = 4\pi n v T_e R^2 \tag{4}$$

For the current Messier 67 calculations, we fitted a stellar density law with a particular case of the generalized Schuster density law (Ninkovic, 1998), which is a simpler version of a King (1962) profile with a finite boundary, and fits as well as a King profile where the radius and central density are known (Bonatto and Bica, 2003). For the velocity dispersion, we used typical observed values (0.5 – 1 km/s) from literature (McNamara and Sanders, 1978; Montgomery *et al.* 1993; Hurley *et al.* 2005), and a reasonable value for $T_e$ is about $4\times10^9$ yr (Hurley *et al.* 2005).

We set the mass of flyby stars to 1 $M_\odot$. For the velocity, we used three different values within the observed range: 0.7 km/s, 0.8 km/s, and 0.9 km/s. Fig. 6 shows both the volumetric stellar density (left panel) and the number of encounters (right panel) plotted versus location of the planetary system on the current cluster. Because the number of encounters depends on the velocity dispersion, we obtained three encounter curves.

From Fig. 6, the amount of stellar encounters on a 200 AU radius is lower than one almost independently of the cluster´s radius (up to 1 pc). This means that M67 seems currently a propitious environment for dynamically cold planetary systems, probably similar to the Solar neighborhood conditions at birth in terms of the average number of encounters.

### 4.3.2. The young Messier 67

In this section, we consider dynamical encounters in the young Messier 67. We employed a Plummer profile to fit the stellar density, and set the values of the total mass (19,000 $M_\odot$) and initial amount of stars (36,000 stars) from the preferred model of Hurley *et al.* (2005) for the young M 67. With these assumptions we constructed a Plummer density law with a core radius of 0.76 pc to simulate this environment. The velocity dispersion is similar in magnitude to those values in star formation regions; therefore, we have taken typical velocities from 1 to 3 km/s (Ida *et al.* 2000).

We chose a Plummer profile because of its simplicity, and because there is no reason to believe that, for example, a King Model should describe the initial state of an open cluster since that model was originally conceived to describe dynamically evolved globular clusters (Hurley *et al.* 2005).

Following the same idea on the current M 67, Fig. 7 shows both the stellar density (left panel) and the number of encounters (right panel) plotted versus location of a planetary system on the young cluster. In this case, we calculated the number of encounters on a 200 AU radius from the host star for a time of $10^7$ yr (while the cluster was young, more massive, and concentrated).

We present here some numerical simulations of the effect on disks by stellar interactions in both the current and the young M 67. Fig. 8 shows an array on which the impact parameter ranges from 40 to 200 AU, the velocity dispersion ranges from 0.5 to 3 km/s, and the mass of the flyby star goes from 0.5 to 1.3 $M_☉$ (the host star is 1 $M_☉$). This array provides a physical idea on how perturbations on disks depend on the flyby mass, velocity, and impact parameter. Red dots (the four upper rows of panels) represent the perturbed disks in the young cluster where typical velocity dispersion values are in the range 2 to 3 km/s, and blue dots (the four lower rows of panels) represent the current cluster with values in the range 0.5 to 1 km/s. As reference, we selected five values for the impact parameter to cover the 200 AU radius where effects of encounters begin to show. Fig. 9 shows a close up to the inner part of the disks.

From Fig. 9, the disk particles eccentricities are lower than 0.2 in all simulations for planets with original semimajor axes lower than 10 AU (approximate location of Saturn in the Solar system), which suggests that Messier 67 was once, and is now, a rather quiet environment in terms of dynamical encounters. If there are planets with semimajor axes within the habitable zone of some stars in this cluster, they may have nearly circular orbits ($e < 0.2$), and therefore they may be potentially habitable (see Section 2).

### 4.3.3. Oort Clouds in the Open Cluster Messier 67

In this Galactic environment, velocity dispersion is relatively low (1-3 km/s). Our calculations suggest that Oort clouds are stripped readily by stellar encounters, similarly to the putative Solar birth cluster. Several other studies have noted that the encounter to produce the orbital parameters of Sedna and heat the Kuiper belt must have occurred rapidly and then the Sun must have left the high density environment before totally forming its Oort cloud (Levison *et al.* 2004), otherwise the Oort cloud would have been stripped out in one relatively close encounter.

To estimate the effect of a flyby on a particle system that is weakly bounded, like an Oort cloud, we show in Fig. 16 the impact parameter times the velocity dispersion divided by the flyby mass. Oort clouds in the young, and in the current, Messier 67 would be stripped out. Even if the Oort cloud had taken $10^9$ years to form, the cluster is still bounded, and the current environment would have destroyed individual clouds. Part of those cometary nuclei might be still bounded in the large potential well of the cluster. On the other hand, if the presence of Oort clouds have a role in habitability, for example, to bring part of the water to planets in the early stages of formation, then this would decrease probabilities for life, as we know it, to occur in this cluster.

### 4.4. Globular Cluster Messier 13

Located in the constellation of Hercules, Messier 13 (M13) is one of the best known globular clusters. It has about $10^5$ stars, spans about 25 pc radius, lies over 6 kpc distance, and is 1.2 × $10^{10}$ yr old. In November 1974, the Arecibo radiotelescope sent a message to hypothetical extraterrestrials in M13. Because of the high stellar density of old stars, M13 was considered a very good target as a planet with intelligent life forms. One problem with regard to life in M13, is its poor metallicity (Soker and Hershenhorn 2007). We studied the dynamical effect of this crowded stellar environment on planetary disks.

The stellar density law is modeled with a King profile. For the velocity dispersion, Lupton *et al.* (1987) provided a value of about 7 km/s for the central part of the cluster. Pryor and Meylan (1993) showed a velocity dispersion list of 56 Galactic globular clusters, and from that list, M13 has a velocity dispersion of 6.62 ± 0.41 km/s. From here, we used three values for our simulations: 3, 5, and 7 km/s.

Using Eq. 4, we calculated the number of encounters in $10^9$ yr, on a 200 AU radius. Because the number of encounters depends on the velocity dispersion, we show three encounter curves. Fig. 10 shows the stellar density (left frame) and encounter curves (right frame).

Fig. 10 shows that planetary systems in the inner regions of Messier 13 are subjected to a high number of close stellar encounters. Between 0 and 5 pc from the center of the cluster, the number of encounters goes from 10 to almost 1000. This Galactic region is a hostile environment, in terms of stellar dynamics, for keeping planets with low orbital eccentricities. In the outskirts of this globular cluster, planetary systems located between 10 and 13 pc sustain approximately one stellar encounter. According to Pryor and Meylan (1993), the radius of Messier 13 is about 25 pc; high orbital angular momentum stars (avoiding the inner parts of the cluster) in external regions of the cluster, from 10 pc to the end of it, could be interesting to search for dynamically cold planetary systems.

We computed eccentricities after a flyby interaction. Our experiments included miss distances of 3, 9, 15, and 45 AU to simulate an impact in the inner region where the closest encounters occur. Then to simulate a flyby impact in the outer zone of the cluster, we set this approach distance to 195 AU. These simulations were modeled by using a 5 km/s velocity dispersion. Fig. 11 shows the resultant eccentricities after the interaction.

In terms of habitability, these results indicate that the central region in Messier 13 is a hostile environment for hosting stable planetary systems. Planets in such a region would be subjected to an enormous amount of stellar encounters. Just one close encounter is able to generate high eccentricities, enough to send potentially habitable planets outside the habitable zone (Section 2). External regions of the cluster, on the other hand, might be interesting because they are unlikely to host close encounters (for the fraction of stars with high orbital angular momentum that will not approach the central parts); a 45 AU typical flyby interaction in such regions cannot produce high eccentricities on planets located in the habitable zone.

It is worth mentioning, however, that in terms of habitability in globular clusters the dynamical complications are just one of the problems. For example, another problem is the lack of metals (compounds heavier than He) in the chemical composition of the cluster, elements that are fundamental to form habitable (rocky) planets and life.

#### 4.4.1. Oort Clouds in the Globular Cluster Messier 13

Globular clusters are one of the most hostile environments in the Galaxy with regard to stable planets around stars. The effect on the weakly bounded Oort clouds is expected to be dramatic. As seen in Fig. 16, Oort clouds would be stripped out by a flyby interaction at almost every radius in the cluster, even without considering that a large percentage of the stellar orbits approach the cluster center.

### 4.5. Galactic Bulge-Bar

For this Galactic environment, we employed the stellar density profile proposed by Freudenreich (1998), who derived a model for the Galactic bar and for the old stellar disk from the COBE/DIRBE (Cosmic Background Explorer/ Diffuse Infrared Background Experiment) observations. He found that the best fit to the bar density goes like the $sech^2$ of the semimajor axis (Fig. 12, left panel). We have included the contribution to the density of the Galactic disk, using the potential of the disk in Allen & Santillán (1991) that fit a Miyamoto-Nagai disk to the Milky Way. For velocity dispersion, Tremaine *et al.* (2002) reported measurements from 0.085 to 1284 pc (empty diamonds in Fig. 12, middle panel). We fit these values to an analytical function (continuous line in the same figure). We finally combine this with the density profile (Fig. 12, middle panel) to obtain the number of stellar encounters and the impact parameters, calculated for a time of 4.5 Gyr. These plots in Fig. 12 represent then the stellar density, velocity dispersion, and the number of encounters within a 200 AU radius, during 4.5 Gyr, all plotted versus position on the bulge-bar, at z=0 (midplane disk), and along the semimajor axis of the bar. The number of encounters in this region is enormous, ranging from about 10 at 800 pc from the Galactic center to about 300 closer to the center of the Galaxy.

To run simulations, we selected 4 different positions from the Galactic center (at z=0): 3.6 pc, 168.5 pc, 520 pc, and 1300 pc. Fig. 13 shows eccentricities and inclinations plotted versus semimajor axes; columns indicate the position of the planetary system on the bulge-bar.

In Fig. 13, orbital parameters show maximum eccentricities smaller than 0.6 for all positions in the cluster, and inclinations are lower than 30°. In this environment, the combined effect of a relative low density and a high velocity dispersion (100 km/s) makes the effect of one stellar interaction with a disk small. This is especially true for regions beyond a galactocentric radius of 1 kpc, approximately, where encounters become rare. Objects in this region, with semimajor axes within the habitable zone would have nearly circular orbits. However, for regions within 1 kpc, the number of encounters may be large enough to compromise the stability of planetary disks. In an ongoing study, we are studying the effects of multiple encounters in crowded regions like this one.

### 4.5.1. Oort Clouds in the Galactic Bulge-Bar

In this environment, the amount of encounters within a 200 AU radius, over 15 Gyr (see Fig. 12), is very large inside a galactocentric radius of about 1 kpc, and although the stellar velocity dispersion is much higher than in the previous Galactic regions (reducing the destructive effect of the encounters), the encounters are so frequent that the probability of Oort clouds to survive is very low. For the external region, beyond approximately 1 kpc, galactocentric radius, the probability of survival is much higher; in the inner region (from approximately 5 to 100 pc), velocity dispersions are between 90 km/s and 130 km/s, and according to Fig. 16, it is necessary to have a 40 AU (or less) encounter to destroy Oort clouds, which is very likely in this region.

## 4.6. Galactic Nucleus

The Galactic nucleus is characterized by a dynamically violent environment. This region possesses high stellar density and velocity dispersion and an unusual star formation. To simulate stellar encounters, we used the typical stellar density of the innermost region (1 pc from the Galactic center), fitted with a power law profile (Genzel *et al.* 2003), centered in the super massive black hole.

The velocity dispersion of the star cluster follows a Keplerian law from 0.1 arcsec to 20 arcsec (0.8 pc) (Genzel *et al.* 2000) due to the presence of a massive compact central object (Genzel *et al.* 1996, 1997, 2000; Ekcart and Genzel, 1996, 1997; Ghez *et al.* 1998). The stellar density law of Genzel *et al.* (2003) is well fitted until 2 pc. Fig. 14 shows the stellar density law, the velocity dispersion, and the amount of stellar encounters on a 200 AU cross section over 1 Gyr (the approximate time for life to occur on Earth).

In this environment, the number of stellar encounters on a 200 AU cross section is enormous at any position along of this star cluster; in fact, there are even good probabilities of collisions between stars. Giant stars could be destroyed (Davies *et al.* 1991; Rasio and Shapiro, 1990; Alexander, 1999) and exotic stars may be formed (Thorne and Zytkow, 1975). Thus, apart from other hostile factors with regard to the development of life in this environment, such as UV radiation, X rays, among others, stellar dynamics would make the development of life and even the survival of planetary systems impossible (Lineweaver *et al.* 2004).

In the Galactic nucleus, velocity dispersions are higher than in the Galactic Bulge-Bar, and as we discussed in section 4.5, effects of a flyby interaction with high velocity may not be high enough to disrupt drastically planetary disks. In Fig. 15, we show a simulation with an encounter velocity 170 km/s at 50 AU, which simulates a typical condition in this Galactic region. Although apparently the effect of one typical encounter might not be strong enough to severely affect a disk, the amount of interactions is high enough to destroy planetary systems on any position along the Galactic nucleus.

### 4.6.1. Oort Clouds in the Galactic Nucleus.

Although this environment has the largest stellar velocity dispersions of the entire Galaxy, which diminishes the effect of individual interactions, along with the extreme density, the nuclear cluster produces a colossal quantity of close encounters. Oort clouds would not survive in this hostile environment.

According to Fig. 16, Oort clouds located up to 0.8 pc from the Galactic center, where the velocity dispersion is about 50-100 km/s, would be destroyed if they were subjected to a 80 AU (or less) encounter. Planetary disks and Oort clouds would sustain an enormous amount of interactions for time periods comparable to the main sequence lifetime of a star ($10^9$ yr), which would make it impossible to preserve any kind of debris structure around stars.

5. Conclusions

In this study, we examined the effect of stellar interactions in different Galactic environments on a 100 AU planetary (debris disk: planets and cometary nuclei) disk. We found that some of these environments generate extreme orbital parameters in planetary disks, which would diminish or even destroy the possibility of planetary habitability. This approach reveals the most likely Galactic environments, from the dynamical point of view, where stars might form and maintain their planets and cometary reservoirs, which are in turn related to the possibility of life within such systems.

In the current Solar neighborhood, Gliese 710 will be the closest star to the Sun, 1.36 Myr in the future with a minimum distance of 0.34 pc and a velocity of 14 km/s. This encounter will lead to direct interactions between the star and the Oort cloud. However, the approach distance is not small enough to disrupt the modeled 100 AU disk, and it will not even produce an important effect on the global structure of the Oort Cloud.

For the birth cloud of the Sun, a flyby interaction is the best known mechanism to have produced some observed properties of the Kuiper belt, in particular, by encounters with velocities between 1-2 km/s and a close-approach distance between 100-150 AU. We can reproduce eccentricities from 0-0.1 on a semimajor axis interval 0-40 AU, from 0 to 1 on a semimajor axis interval 40-65 AU, and from 65 to 100 AU where the most particles are dynamically evaporated.

Regarding the open cluster Messier 67, we found that, by modeling its current and young epochs, the number of stellar encounters on a 200 AU cross section disk is not high. For the current M67, it is lower than one. Numerical simulations showed unperturbed inner regions (0-10 AU) on the 100 AU disks. For the modeled current and young cluster, we observed objects with eccentricities lower than 0.2, which suggests that Messier 67 could be an interesting environment to search dynamically for cold planetary systems with planets on near circular orbits.

According to our simulations, a planetary system located in the inner region of Messier 13 (from the center to approximately 5 pc) is subjected to a high number of close stellar encounters over its lifetime. This could suggest that the central region is a hostile environment in which to search for cold dynamically planetary systems. External regions of the cluster (from approximately 10 pc and thereafter) could be interesting because it is unlikely to have close encounters; one flyby interaction cannot produce high eccentricities on planets located in the habitable zone. It is worth mentioning that another problem is the lack of metals in the chemical composition of globular clusters, which may hinder the occurrence of planets.

In the Galactic bulge-bar, the effect of the high velocity dispersion (100-150 km/s) diminishes the probability of destroying a disk in one single encounter, however, the large density produces a large quantity of encounters, making a hostile environment to keep planetary orbits unperturbed around their host stars in the region inside of 1 kpc approximately. This result represents a first approach in dynamical terms, other factors may affect the development and evolution of life in the central Galactic regions.

Regarding the Galactic nucleus, the number of encounters on the modeled 200 AU cross section is high enough to destroy completely disks. We have calculated these encounters during $10^9$ yr (reasonable time for the development and evolution on life as it is estimated for life on Earth). The Galactic nucleus is an inappropriate environment for searching cold dynamically planetary systems.

## Acknowledgments

J.J. And B. P. thank project UNAM through grant PAPIIT IN-110711. A.S. Acknowledges the support from the Universidad Nacional Autónoma de México grant DGAPA PAPIIT IA101312 and CONACYT grant No. 128228.

## Author Disclosure Statement

No competing financial interests exist.

Gilmore, G., Grebel, E. K., Helmi, A., Navarro, J. F., Anguiano, B., Boeche, C., Burton, D., Cass, P., Dawe, J., Fiegert, K., Hartley, M., Russell, K., Veltz, L., Bailin, J., Binney, J., Bland-Hawthorn, J., Brown, A., Dehnen, W., Evans, N. W., Re Fiorentin, P., Fiorucci, M., Gerhard, O., Gibson, B., Kelz, A., Kujken, K., Matijevi, G., Minchev, I., Parker, Q. A., Pearrubia, J., Quillen, A., Read, M. A., Reid, W., Roeser, S., Ruchti, G., Scholz, R.-D., Smith, M. C., Sordo, R., Tolstoi, E., Tomasella, L., Vidrih, S., Wylie-de Boer, E. (2008) The Radial Velocity Experiment (RAVE): Second Data Release. *Astron. J.* 136: 421-451.

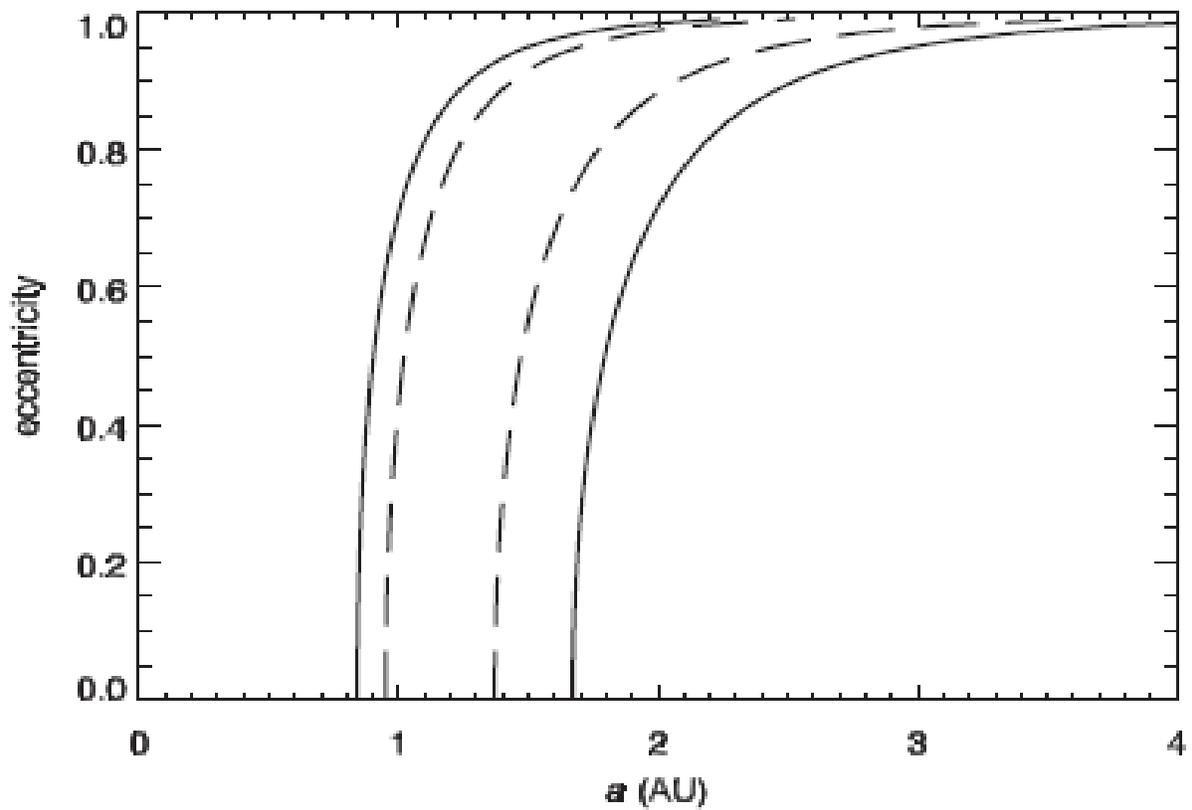

Figure 1. Boundaries of the eccentric habitable zone calculated from Eq. 8 and 9 by Barnes et al. (2008). The inner boundaries correspond to the runaway greenhouse limit (solid line) and the water loss limit (dashed line) as calculated by Kasting et al. (1993). The outer boundaries correspond to the first $CO_2$ condensation limit (dashed line) and the maximum greenhouse limit (solid line) as calculated by Kasting et al. (1993).

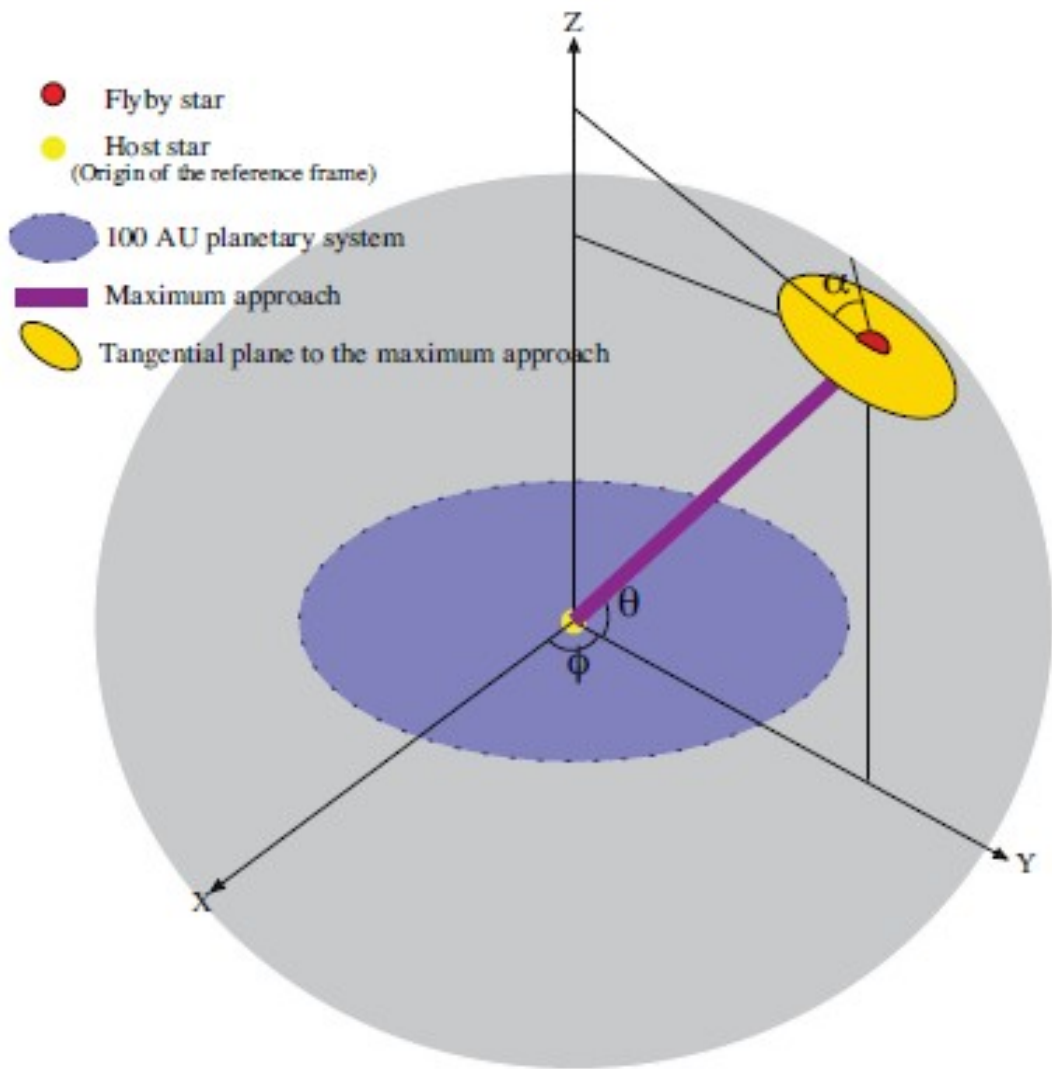

Figure 2. Schematic figure of a stellar encounter on a planetary system by a flyby star. The disk is simulated with test particles. The initial conditions are particles on circular orbits with zero inclination.

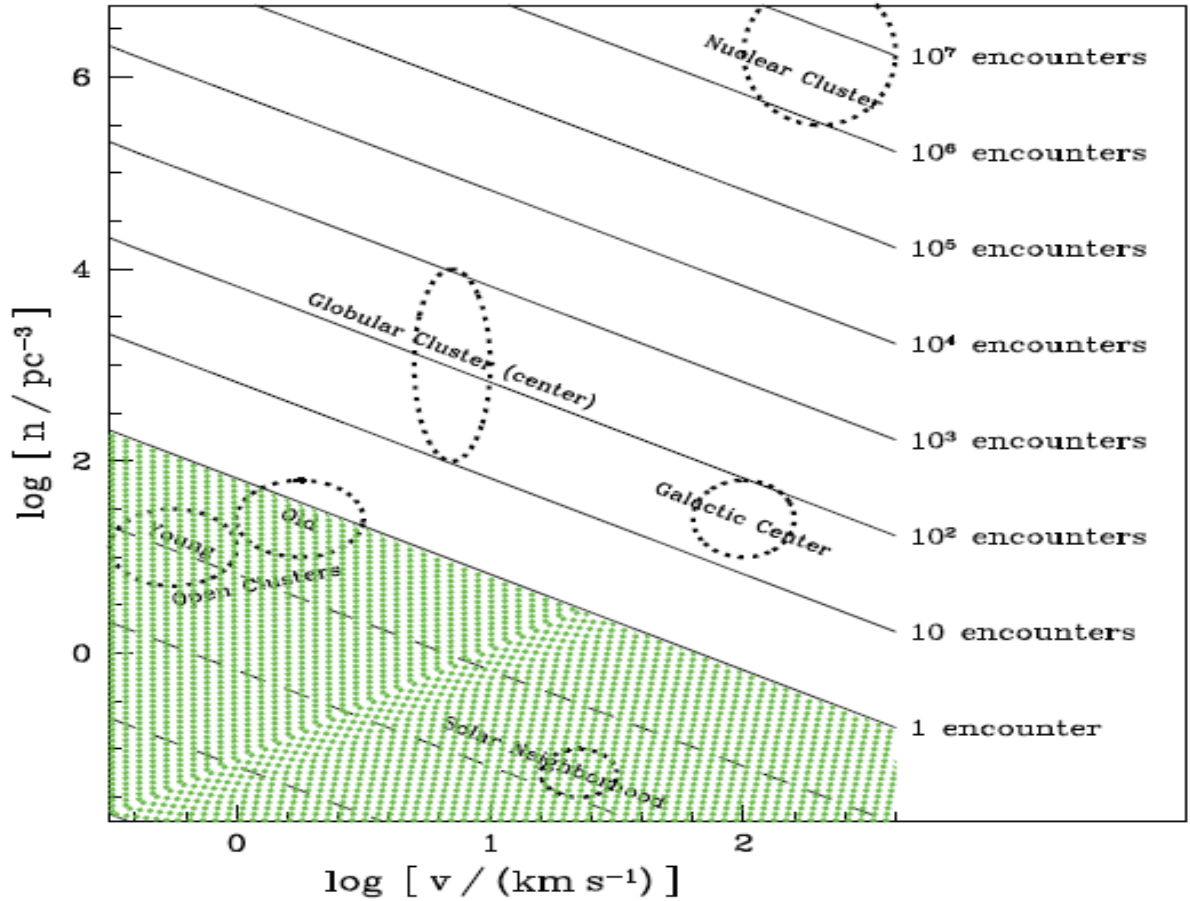

Figure 3. Log-log diagram of density vs. velocity dispersion in different Galactic environments, marked with an elliptical region that approximates typical values from literature. Straight lines represent the number of encounters (eq. 4), given a density and velocity dispersion for a total integration time $T_e$ of 5 Gyr (for all environments). All environments included have existed for the integration times we employed (the most of them even more), except for young clusters (they live bounded about $10^8$ years); however, this environment is so rarified that the number of encounters is almost the same in the total integration time $T_e$ employed. The green shadow covers the galactic regions where less than one stellar encounter occurred in its history; these regions are potentially habitable from the stellar encounters dynamics point of view.

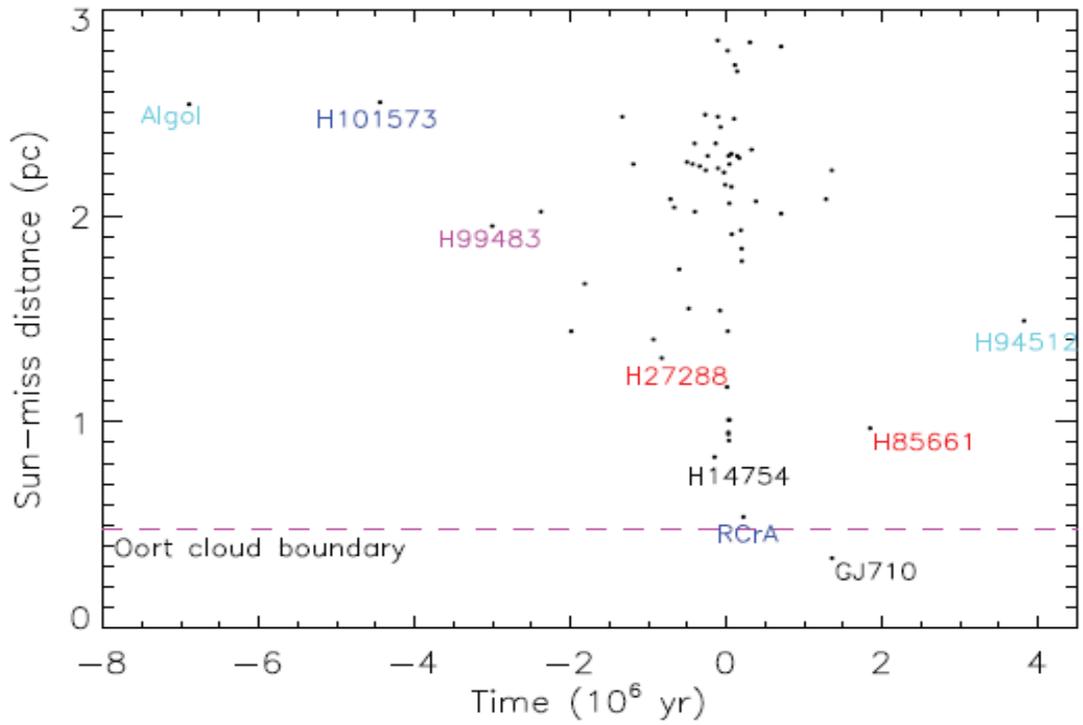

Figure 4. Miss distances vs. time. We only show stars with impact parameter (miss distance) < 3 pc. The Oort cloud boundary (0.48 pc) is marked as reference (dotted line). The closest approach will be with the star Gliese 710 with an impact parameter of 0.34 pc, 1.36 Myr in the future.

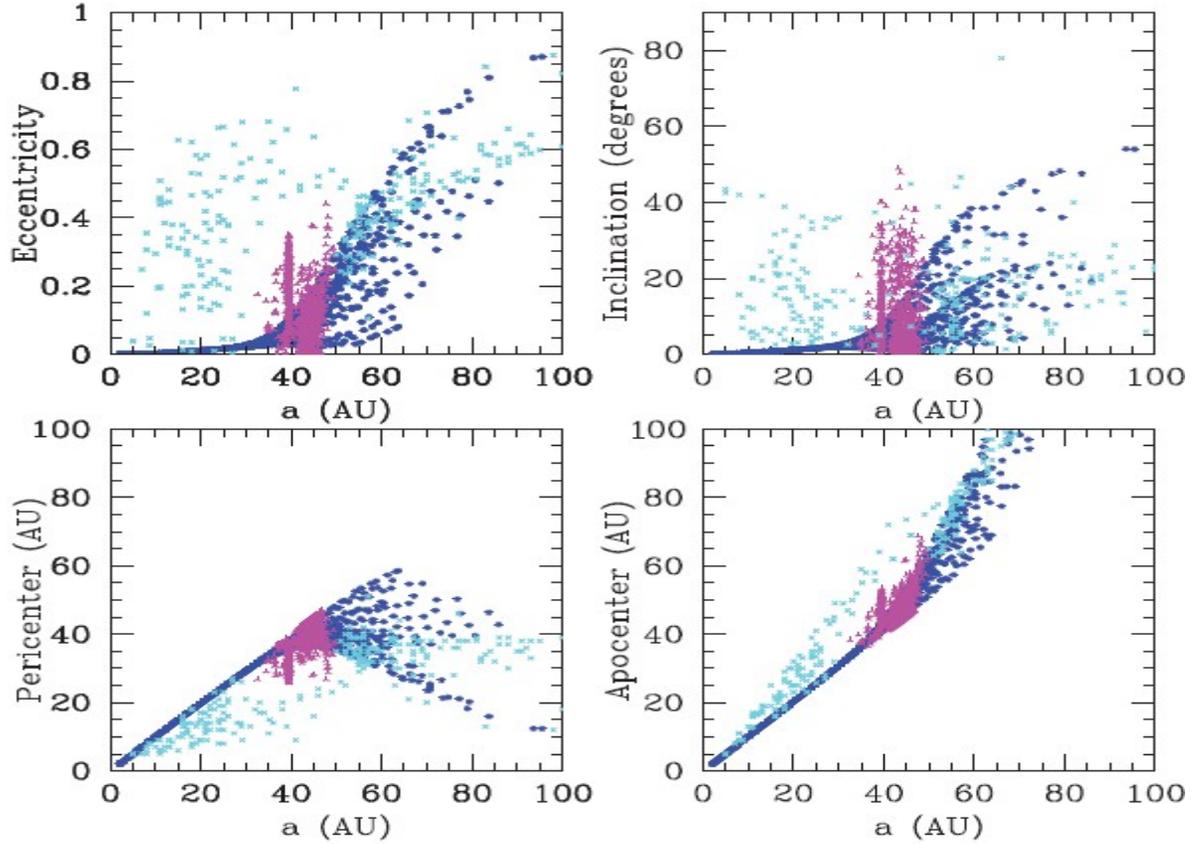

Figure 5. Orbital parameters after a stellar encounter: eccentricities (upper left), inclinations (upper right), pericenters (lower left), and apocenters (lower right). Resonant objects and classic Kuiper objects are included (pink triangles), and scattered objects and Centaurs at radii less than 30 AU (cyan crosses); blue dots represent our numerical results. This array shows an experiment with a 150 AU miss distance and a 1 km/s flyby velocity.

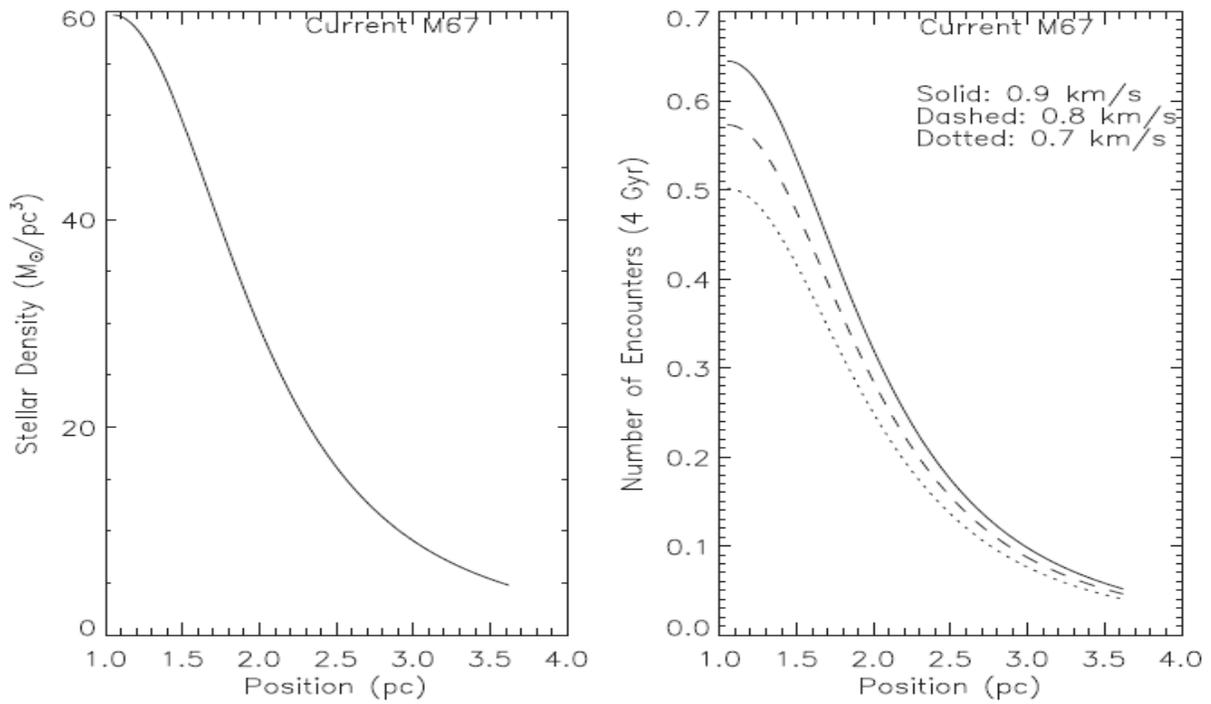

Figure 6. Stellar density and number of encounters (to distances between stars less than 200 AU) plotted versus radial position of the 100 AU planetary system on the current Messier 67. These calculations were done for a time, $T_e = 10^9$ yr. For reasonable velocity values (0.9, 0.8 and 0.7 km/s), the amount of encounters on the 200 AU cross section is lower than one.

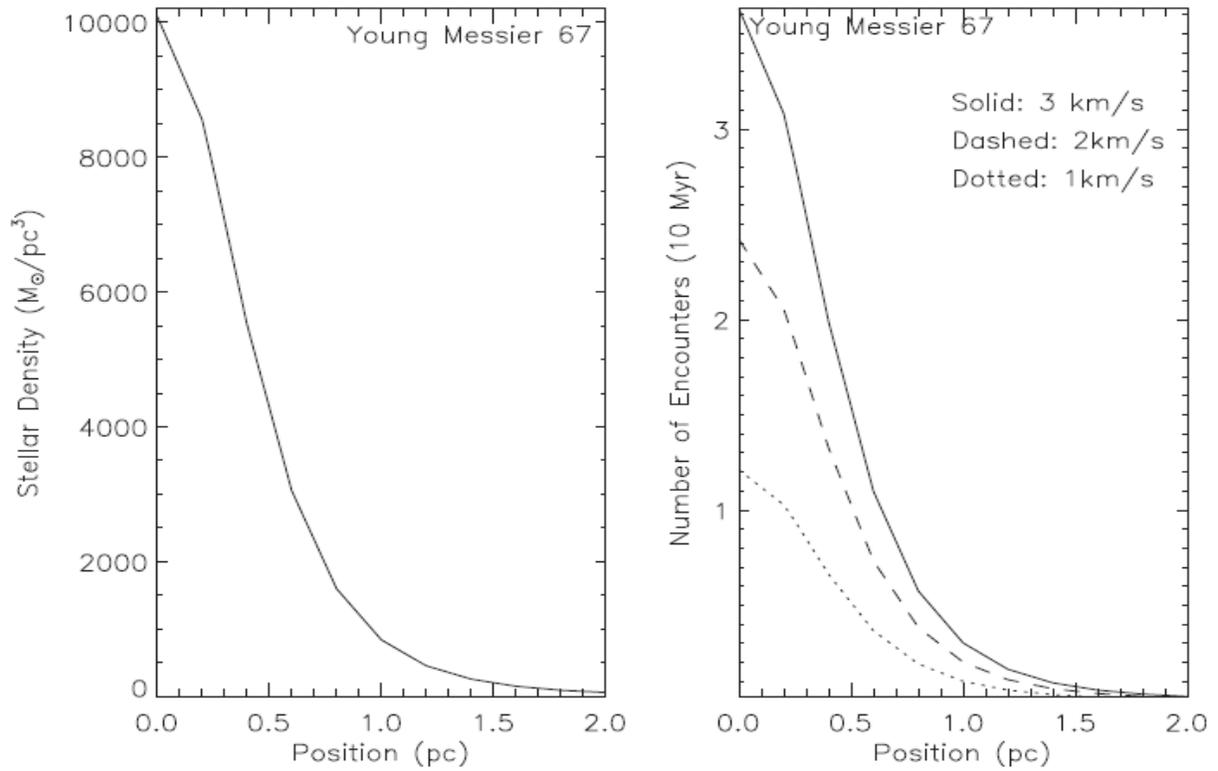

Figure 7. Stellar density and number of stellar encounters plotted versus radial position of a 100 AU planetary system on the young Messier 67. These calculations were done for a time, Te = $10^7$ yr. The number of encounters in a 200 AU cross section are approximately lower than three.

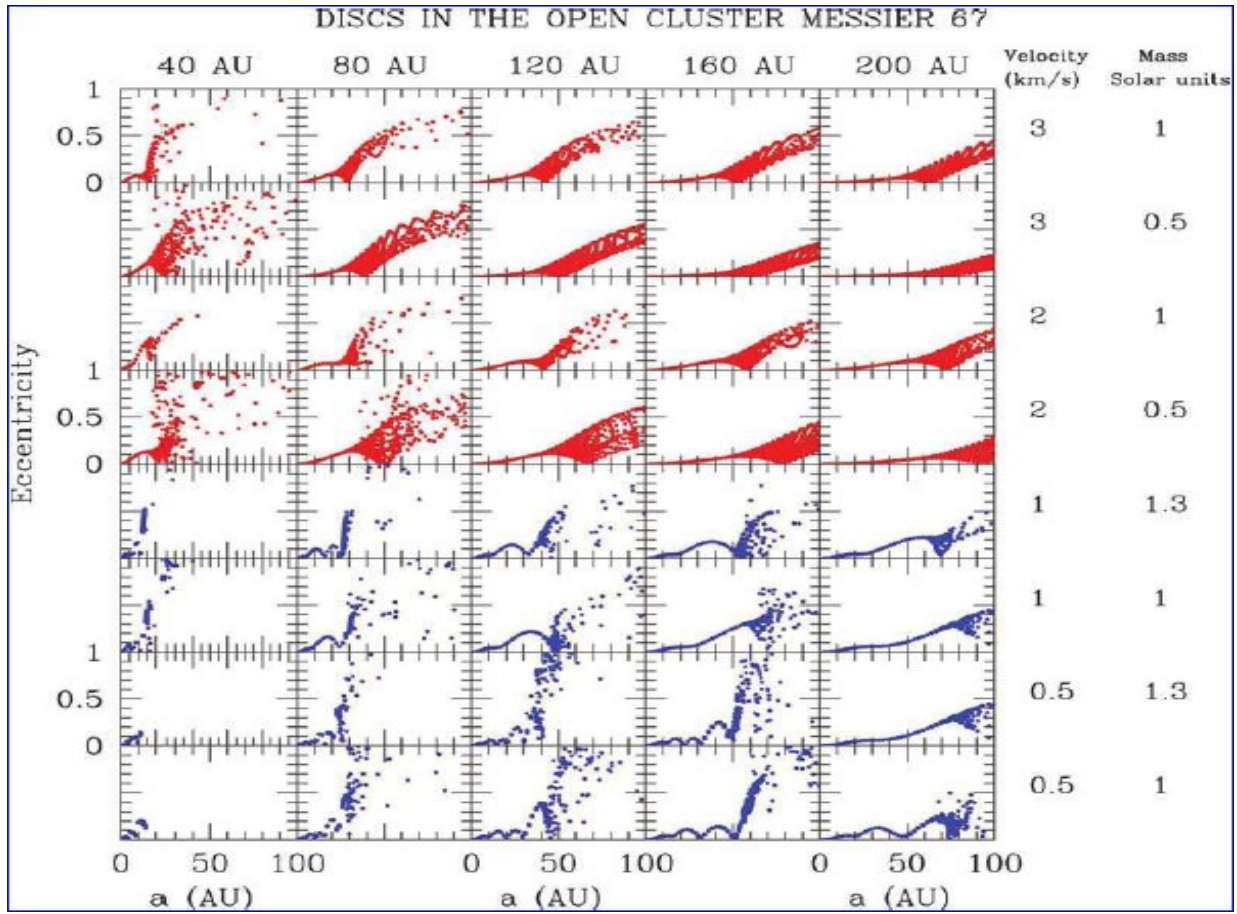

Figure 8. Disks in the cluster Messier 67. Red dots (four upper rows) represent the young cluster for which typical velocity dispersion values are 2-3 km/s. The blue dots (the four lower rows) simulate the current cluster with reasonable velocity dispersion values between 0.5 and 1 km/s. The array shows the results for stellar masses between 0.5 and 1.3 M☉. The columns indicate the flyby maximum approach distance to the 100 AU disks.

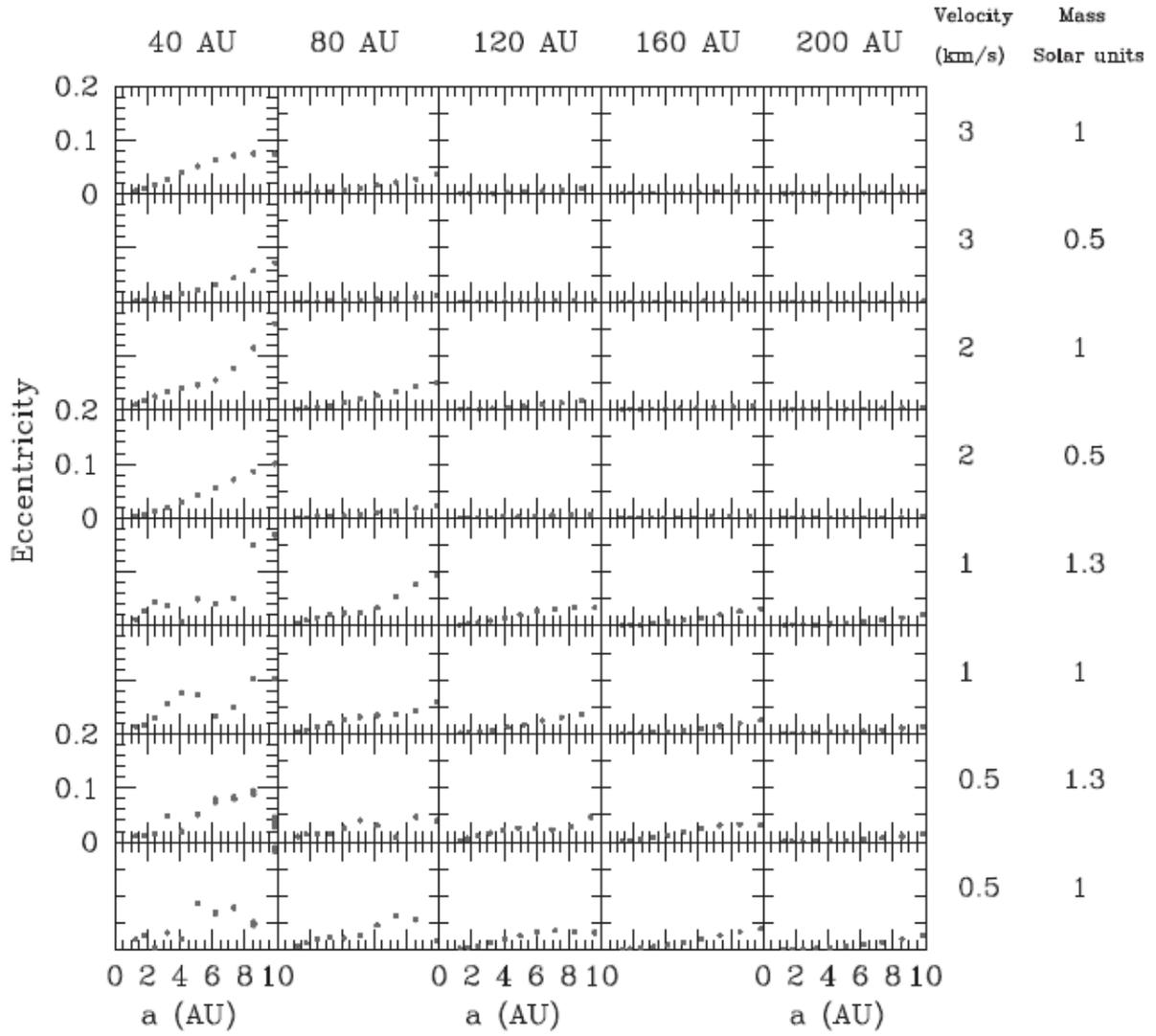

Figure 9. Close up of Fig. 10 from 0 AU to 10 AU, eccentricities are lower than 0.2.

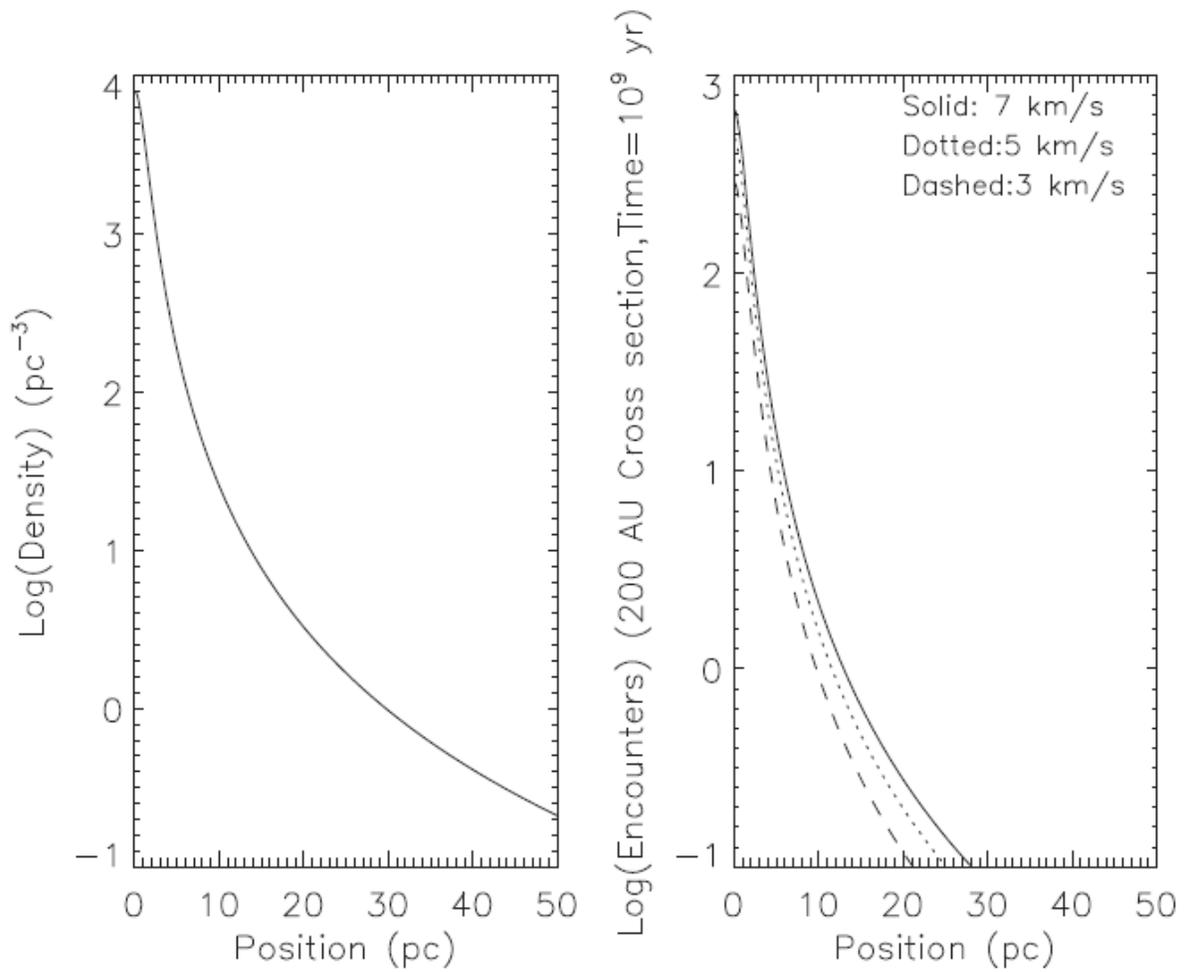

Figure 10. Volumetric stellar density (left) and number of encounters (right) in $10^9$ yr, plotted versus position of planetary systems on the globular cluster Messier 13. The right frame shows results for 3 different velocity dispersion values.

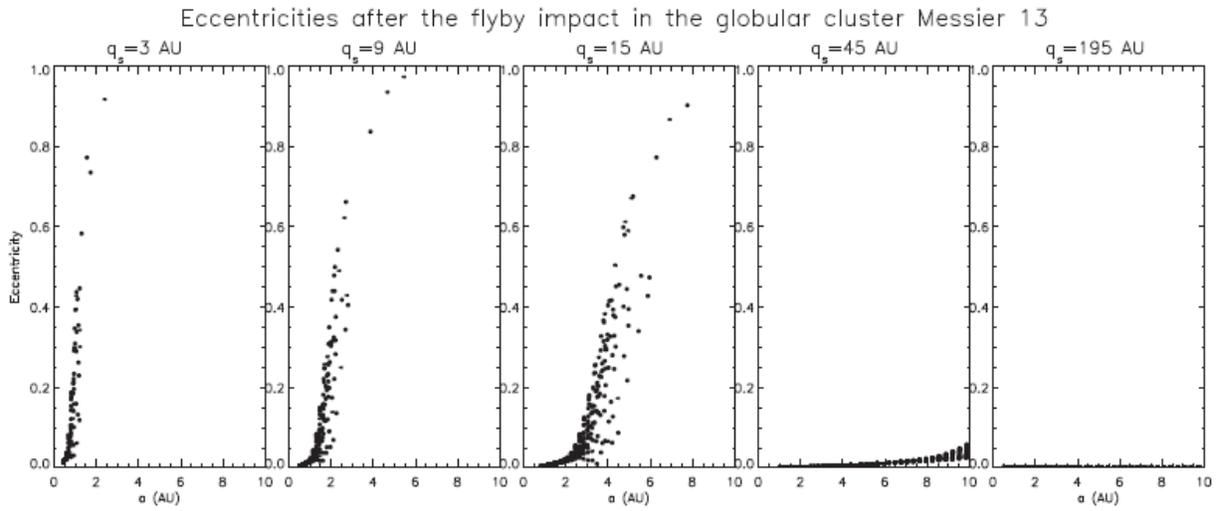

Figure 11. Disk particle eccentricity after the flyby impact, simulating close encounters in the globular cluster Messier 13. These simulations were modeled by using a 5 km/s velocity dispersion. qs means the miss distance for these experiments.

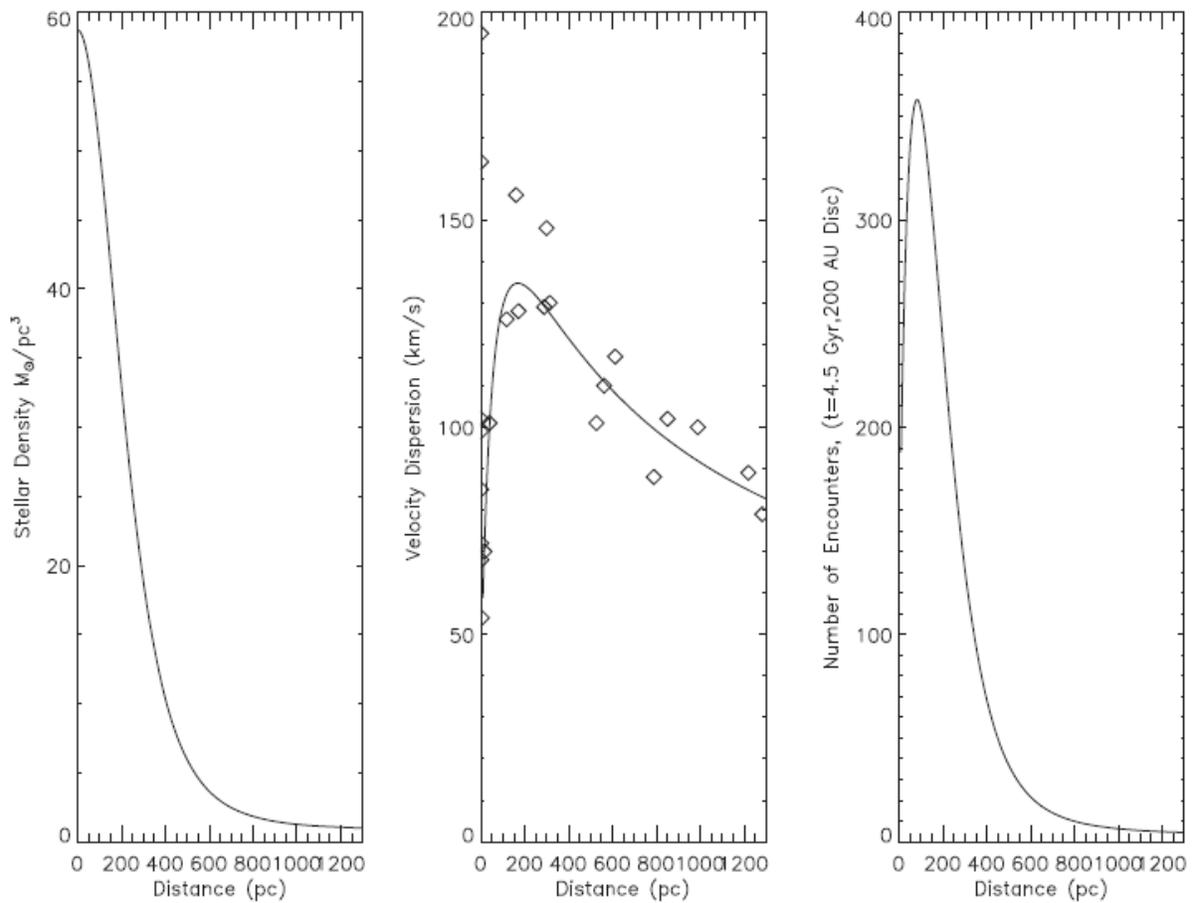

Figure 12. Curves of stellar density, velocity dispersion (a fit from observations -marked with empty diamonds-), and number of stellar encounters on a 200 AU disk in 4.5 Gyr plotted versus position on the Galactic Bulge-Bar.

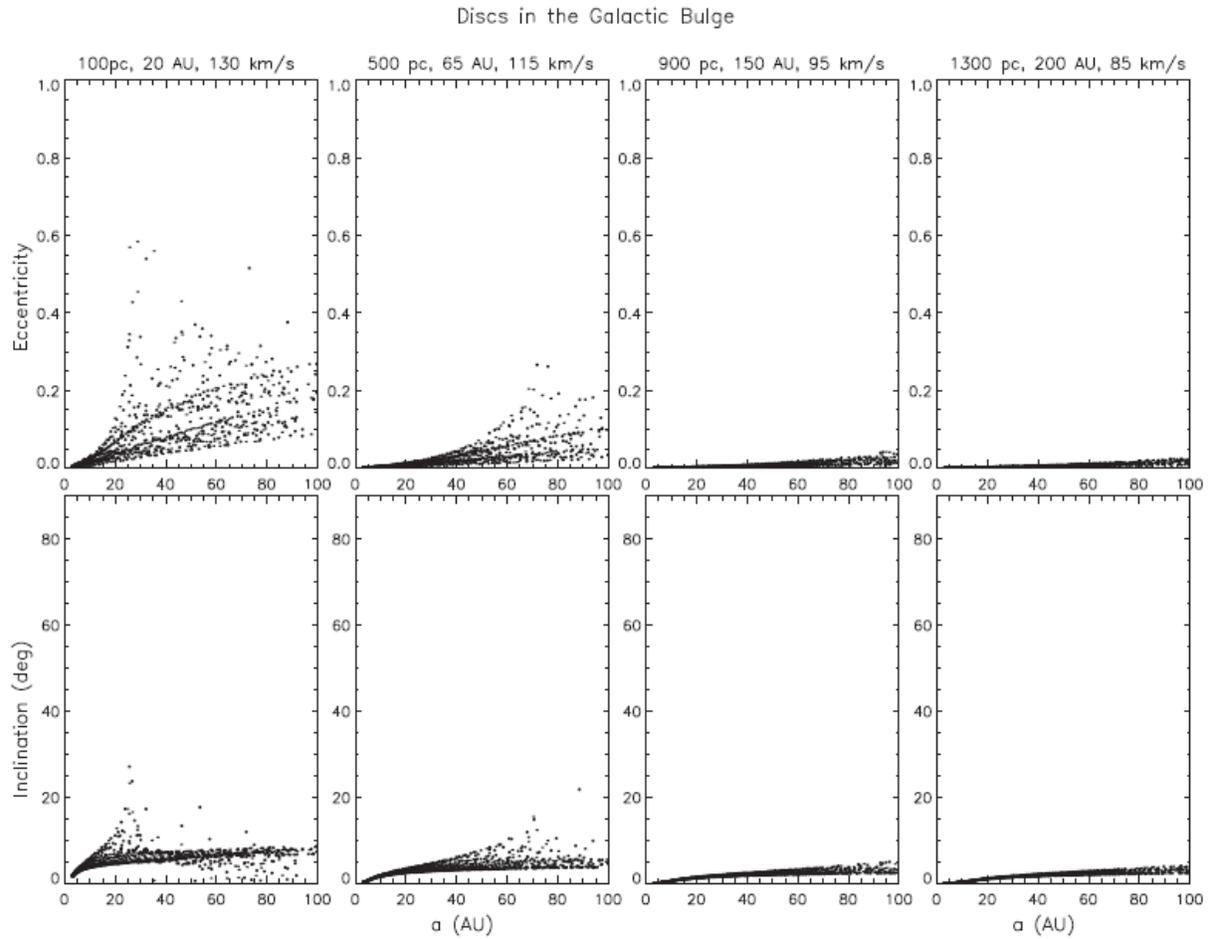

Figure 13. Eccentricities and inclinations plotted versus semimajor axes. As can be seen, eccentricities and inclinations are lower than 0.1 and 10, respectively; this is a reasonable result because the velocity dispersion is high in the Galactic Bulge-Bar.

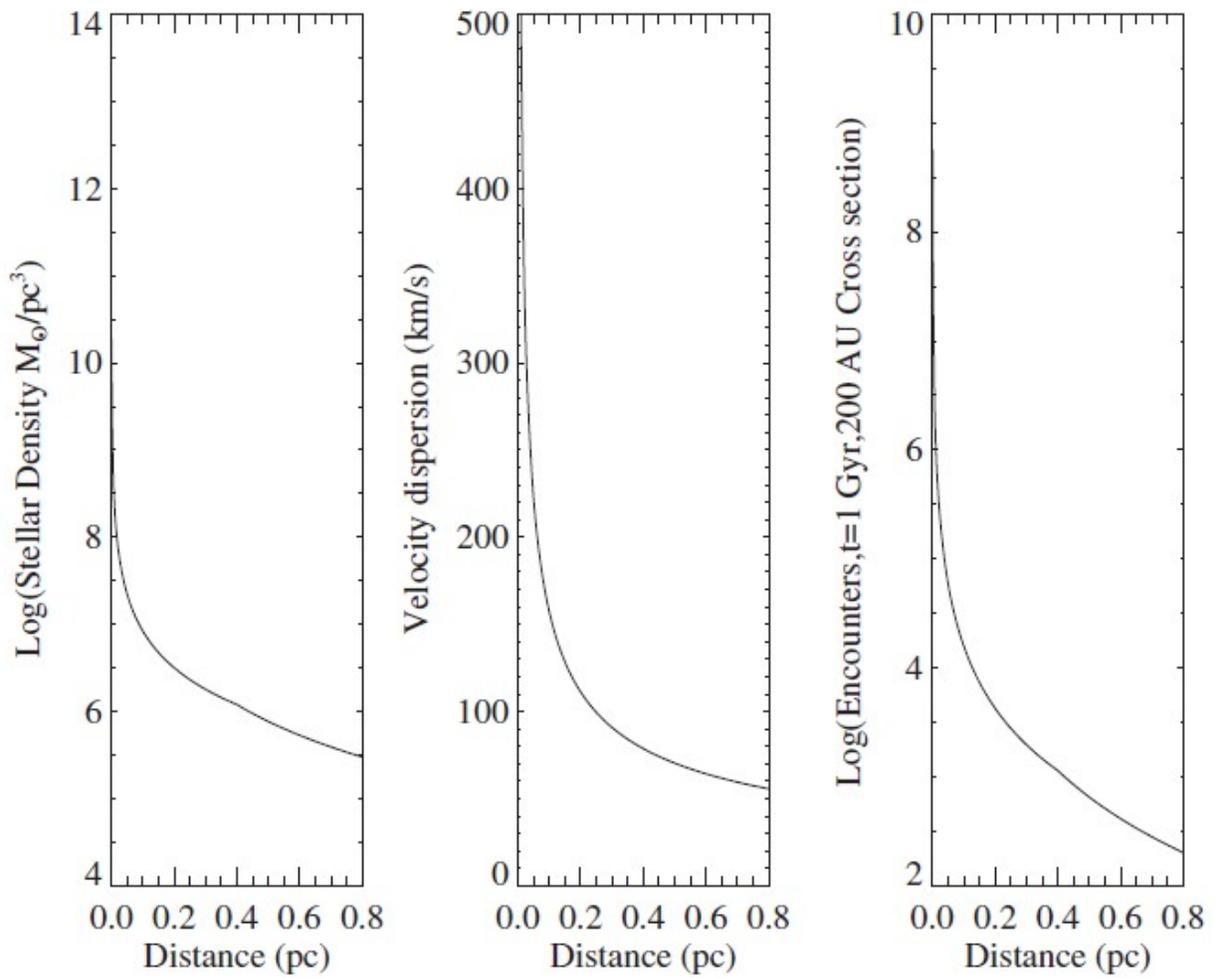

Figure 14. Stellar density, velocity dispersion, and number of stellar encounters on a 200 AU cross section disk calculated for a time of $10^9$ yr.

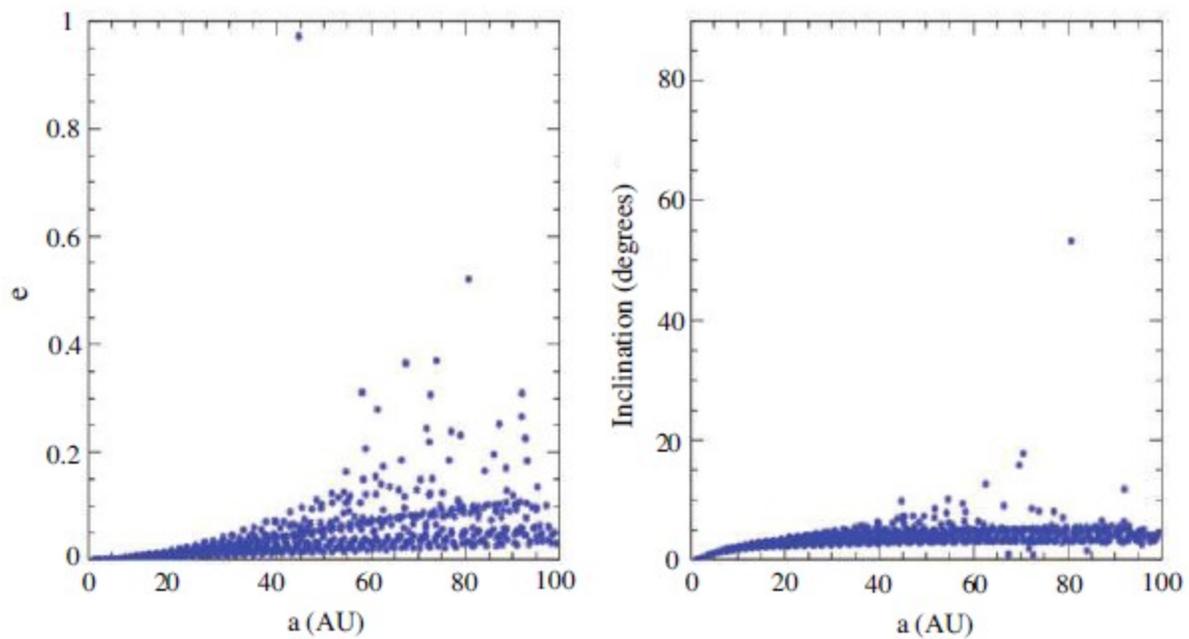

Figure 15. Eccentricities and inclinations plotted versus semimajor axis for an impact parameter of 50 AU with a velocity of 170 km/s, representing a typical condition in the Galactic center.

| q=40 AU | q=80 AU | q=120 AU | q=160 AU | q=200 AU | v (km/s) | Mass Solar units |
|---|---|---|---|---|---|---|
| | | | | | 500 | 1.0 |
| | | | | | 400 | 1.0 |
| | | | | | 300 | 1.0 |
| | | | | | 200 | 1.0 |
| ▓ | | | | | 130 | 1.0 |
| ▓ | | | | | 110 | 1.0 |
| ▓ | | | | | 100 | 1.0 |
| ▓ | | | | | 90 | 1.0 |
| ▓ | ▓ | | | | 70 | 1.0 |
| ▓ | ▓ | | | | 50 | 1.0 |
| ▓ | ▓ | ▓ | ▓ | ▓ | 7 | 1.0 |
| ▓ | ▓ | ▓ | ▓ | ▓ | 5 | 1.0 |
| ▓ | ▓ | ▓ | ▓ | ▓ | 3 | 1.0 |
| ▓ | ▓ | ▓ | ▓ | ▓ | 3 | 0.5 |
| ▓ | ▓ | ▓ | ▓ | ▓ | 2 | 1.0 |
| ▓ | ▓ | ▓ | ▓ | ▓ | 2 | 0.5 |
| ▓ | ▓ | ▓ | ▓ | ▓ | 1 | 1.3 |
| ▓ | ▓ | ▓ | ▓ | ▓ | 1 | 1.0 |
| ▓ | ▓ | ▓ | ▓ | ▓ | 0.5 | 1.3 |
| ▓ | ▓ | ▓ | ▓ | ▓ | 0.5 | 1.0 |

Figure 16. Velocity times impact parameter (miss distance) divided by the flyby mass, according to the equation 3; shaded zones show conditions to strip out the exo-Oort clouds. The mass in the last column represents the flyby mass. The letter q means the impact parameter and the letter v corresponds to the velocity dispersion. This Figure includes several values of miss distances, velocities, and flyby masses. These values cover enough data to simulate parameters of the modeled different Galactic environments.